\documentclass[12pt,a4paper]{article}
\usepackage{amsfonts}

\usepackage{sw20lart}

\input{tcilatex}
\begin{document}

\author{Mekhfi.M \\
\\
{\normalsize Int' Centre for theoretical Physics ,Trieste 34100 \textbf{ITALY%
}} {}\thanks{{{{{\protect\footnotesize Mailing address:Laboratoire de
physique math\'{e}matique ,D\'{e}partement de physique Universit\'{e}
Es-senia,31100 oran-ALGERIE}}}}}}
\title{The $M_{L}(z);C_{L}(z);W_{L}(z)$\ associated Laguerre Polynomials}
\date{2000 June 16}
\maketitle

\begin{abstract}
In a previous paper we deformed Hermite polynomials to three associated
polynomials .Here we apply the same deformation to Laguerre
polynomi\nolinebreak als .



{\footnotesize Mekhfi @hotmail.com}

\newpage{}
\end{abstract}

\smallskip In recent papers we applied a deformation mechanism on Bessel and
Neumann functions of integer orders $J_{n}(z)$ , $N_{n}(z)$ \cite
{chamseddine} and Hermite polynomials $H_{n}(z)$ \cite{mohsine} to generate
respectively Bessel and Neumann functions of real orders $J_{n+\lambda
}(z),N_{n+\lambda }(z)$ ($\lambda $ real ) and Hermite associated
polynomials $M_{n\beta ,H}^{s}(z);C_{n\beta ,H}^{s}(z);W_{n\beta ,H}^{s}(z)$%
\textbf{\ .}The structure underlying the deformation of Hermite polynomials
has some characteristics.

\begin{itemize}
\item  The associated polynomials come in triplicate and form the sequence

$H_{n}(z)\rightarrow M_{n\alpha ,H}^{s}(z)\Rightarrow C_{n\alpha
,H}^{s}(z)\leftarrow W_{n\alpha ,H}^{s}(z)\Leftarrow H_{n}(z)$

\item  When we try to define a measure $\frak{D}_{s}$ which ensures
orthogonality of polynomials\thinspace \thinspace \thinspace \thinspace $\,M$
$^{^{\prime }s}$ we find that $\frak{D}_{s}$ is not positive .It is rather a
``charge'' density .Moreover this measure has the form $\frak{D}_{s}\sim 
\frak{D}_{0}H_{s}$ where $\frak{D}_{0}$ is the measure of Hermite polynomials

\item  The differential equation associated to $M$ is inhomogeneous whose
homogeneous part is Hermite polynomials differential equation.The
inhomogeneous term being of the form $\sim $ $s\sum_{m=1}^{n}\alpha ^{m}%
\frac{d^{m}}{dz^{m}}M_{n\alpha ,H}^{s}(z)$
\end{itemize}

The above structure seems to be general and not restricted to Hermite
polynomials.We will see in this paper that the same structure appears when
we deform Laguerre polynomials and that the above three characteristics
remain valid where $H$ is replaced by $L$

\section{The $M_{n\beta }^{s}$\protect\smallskip $(z)$ polynomials}

\subsection{Definition}

We define Polynomials $M_{n\beta \alpha }^{s}$\smallskip $(z)$ as follows%
\footnote{%
we will use the short notation $M,C,W$ without the subscript $L$ , as only
Laguerre deformations are involved in this paper.Also $d_{m}$ will stand for 
$d_{m}=\frac{d^{m}}{dz^{m}}$}

\begin{equation}
M_{n\beta \alpha }^{s}(z)=\exp [s\sum_{m=1}^{\infty }\frac{\alpha ^{m}}{m}%
\frac{d^{m}}{dz^{m}}]\ L_{n\beta }(z)  \label{eq:1}
\end{equation}

with $\alpha =\pm 1\ $ and $L_{n\beta }(z)$ is Laguerre polynomials which
are defined as follows \cite{wissale}

\[
n!L_{n\beta }(z)=\sum_{m=0}^{n}(-1)^{m}\left( 
\begin{array}{c}
n+\beta \\ 
n-m
\end{array}
\right) \frac{z^{m}}{m!} 
\]

Laguerre polynomials obey a functional relation which is at the origin of
their deformation defined at \ref{eq:1}.It is

\begin{equation}
\frac{d}{dz}L_{n\beta }(z)=-L_{(n-1)(\beta +1)}(z)  \label{eq:3}
\end{equation}

We can analyse polynomials $\,$\-$M_{n\beta \alpha }^{s}(z)$ using the
defining equation \ref{eq:1} or equivalently through their generating
function which can also be used to infer the measure associated to them.

\subsection{Generating function}

The Laguerre polynomials generating function $L(z,t,\beta
)=\sum_{n=0}^{\infty }L_{n\beta }(z)t^{n}\,\ \nolinebreak \mid \nolinebreak
t\nolinebreak \mid \nolinebreak \langle \nolinebreak 1$ has a functional
relation

\begin{equation}
\frac{\partial }{\partial z}L(z,t,\beta )=-tL(z,t,\beta +1)  \label{eq:3'}
\end{equation}

This is property \ref{eq:3} expressed in terms of the generating
function.The action of the exponential deformation on $L_{n}(z)$ is not
straightforward (as property \ref{eq:3} involves the variation of both $n$
and $\beta $ ) ,neither for $L(z,t,\beta )$ (as property \ref{eq:3'}
involves the variation of $\beta $ ).We therefore look for another related
function whose deformation is straightforward.The appropriate function is
defined as follows. Put $\beta =m+\lambda $ with $m$ integer and $\lambda $
a real number and define the function $\Phi (z,t,\zeta ,\lambda )$ formally
as

\[
\Phi (z,t,\zeta ,\lambda )=\sum_{m=-\infty }^{\infty }L(z,t,\beta )\zeta
^{\beta } 
\]

where $L(z,t,\beta )$ coincides with the generating function for Laguerre
polynomials when $\func{Re}$ $\beta \rangle -1$ $.$ The exact form of $\Phi
(z,t,\zeta ,\lambda )$ for arbitrary $\beta $ does not matter .This function
has a simple property.It is an eigenstate of $\frac{\partial }{\partial z}$

\[
\frac{\partial }{\partial z}\Phi (z,t,\zeta ,\lambda )=-\frac{t}{\zeta }\Phi
(z,t,\zeta ,\lambda ) 
\]

In fact

\begin{eqnarray*}
\frac{\partial }{\partial z}\Phi (z,t,\zeta ,\lambda ) &=&\sum_{m=-\infty
}^{\infty }\frac{\partial }{\partial z}L(z,t,\beta )\zeta ^{\beta } \\
&=&-t\sum_{m=-\infty }^{\infty }L(z,t,\beta +1)\zeta ^{\beta } \\
&=&-\frac{t}{\zeta }\sum_{m=-\infty }^{\infty }L(z,t,\beta )\zeta ^{\beta }
\\
&=&-\frac{t}{\zeta }\Phi (z,t,\zeta ,\lambda )
\end{eqnarray*}

It is the function $\Phi (z,t,\zeta ,\lambda )$ that is simple to deform as
it is an eigenstate of the derivative operator.Applying the mapping operator
in \ref{eq:1} we get

\begin{eqnarray*}
\exp [s\sum_{m=1}^{\infty }\alpha ^{m}\frac{d_{m}}{m}]\Phi (z,t,\zeta
,\lambda ) &=&\exp [-s\sum_{m=1}^{\infty }\frac{(-1)^{m+1}}{m}(\alpha \frac{t%
}{\zeta })^{m}]\Phi (z,t,\zeta ,\lambda ) \\
&=&\exp [-s\ln (1+\alpha \frac{t}{\zeta })]\Phi (z,t,\zeta ,\lambda ) \\
&=&(1+\alpha \frac{t}{\zeta })^{-s}\Phi (z,t,\zeta ,\lambda )
\end{eqnarray*}

The deformed generating function associated to $L(z,t,\beta )$ is denoted $%
M(z,t,\beta ,\alpha )$ and is by definition

\begin{equation}
(1+\alpha \frac{t}{\zeta })^{-s}\Phi (z,t,\zeta ,\lambda )=\sum_{m=-\infty
}^{\infty }M(z,t,\beta ,\alpha )\zeta ^{\beta }  \label{eq:8}
\end{equation}

Inserting the definition of $\Phi (z,t,\zeta ,\lambda )$ into \ref{eq:8}
together with the expansion

\[
(1+\alpha \frac{t}{\zeta })^{-s}=1-s\alpha \frac{t}{\zeta }+\frac{s(s+1)}{2!}%
\left( \frac{t}{\zeta }\right) ^{2}+\ldots 
\]

we get the result

\begin{eqnarray*}
(1+\alpha \frac{t}{\zeta })^{-s}\Phi (z,t,\zeta ,\lambda )
&=&\sum_{m=-\infty }^{\infty }(1+\alpha \frac{t}{\zeta })^{-s}L(z,t,\beta
)\zeta ^{\beta } \\
&=&\sum_{m=-\infty }^{\infty }L(z,t,\beta )\zeta ^{\beta } \\
&&-s\alpha t\sum_{m=-\infty }^{\infty }L(z,t,\beta )\zeta ^{\beta -1} \\
&&+\frac{s(s+1)}{2!}t^{2}\sum_{m=-\infty }^{\infty }L(z,t,\beta )\zeta
^{\beta -2} \\
&&+\ldots \\
&=&\sum_{m=-\infty }^{\infty }(L(z,t,\beta )-s\alpha tL(z,t,\beta +1) \\
&&+s(s+1)t^{2}L(z,t,\beta +2)+\ldots )\zeta ^{\beta }
\end{eqnarray*}

Comparing with \ref{eq:8} we identify the generating function for $M$
polynomials which reads

\[
M(z,t,\beta )=L(z,t,\beta )-s\alpha tL(z,t,\beta +1)+s(s+1)t^{2}L(z,t,\beta
+2)+\ldots 
\]

To sum up the above series we replace the generating function of Laguerre
polynomials by its expression $L(z,t,\beta )=\frac{e^{\frac{-zt}{1-t}}}{%
(1-t)^{\beta +1}}$ and get

\begin{eqnarray*}
M(z,t,\beta ,\alpha ) &=&\frac{e^{\frac{-zt}{1-t}}}{(1-t)^{\beta +1}}\left(
1-s\alpha t\frac{1}{1-t}+s(s+1)(\frac{\alpha t}{1-t})^{2}+\cdots \right) \\
&=&\frac{e^{\frac{-zt}{1-t}}}{(1-t)^{\beta +1}}\left( \frac{1-t(1-\alpha )}{%
1-t}\right) ^{-s} \\
&=&\frac{e^{\frac{-zt}{1-t}}}{(1-t)^{\beta -s+1}}\left( 1-t(1-\alpha
)\right) ^{-s}
\end{eqnarray*}

Polynomials $M$ ,them ,are by definition

\[
\frac{e^{\frac{-zt}{1-t}}}{(1-t)^{\beta -s+1}}\left( 1-t(1-\alpha )\right)
^{-s}=\sum_{n=0}^{\infty }\frac{M_{n\beta \alpha }^{s}(z)}{n!}t^{n} 
\]

At this point we may distinguish two different cases.The case $\alpha =1$
does not yield a new generating function but the generating function of
Laguerre polynomials shifted by $s$ i.e. $\beta \rightarrow \beta -s$ .This
is however an interessting result as it allows us to connect Laguerre
polynomials with different value \footnote{%
The parameter $s$ can be taken to be real and of any sign in the case $%
\alpha =1$ .We will see that in the more interesting case $\alpha =-1$
,partial-orthogonality will require $s$ to be only integer and positive} of
the parameter $\beta $

\[
L_{n(\beta -s)}(z)=\exp [s\sum_{m=1}^{\infty }\frac{d_{m}}{m}]\ L_{n\beta
}(z) 
\]

The second more interresting case $\alpha =-1$ will lead to a new family of
Laguerre related polynomials .They are obtained via the relation $($denote $%
M_{n\beta (-1)}^{s}(z)=M_{n\beta }^{s}(z))$

\begin{eqnarray}
\frac{e^{\frac{-zt}{1-t}}}{(1-t)^{\beta -s+1}}\left( 1-2t)\right) ^{-s}
&=&\sum_{n=0}^{\infty }\frac{M_{n\beta }^{s}(z)}{n!}t^{n}  \label{eq:13"} \\
M_{n\beta }^{s}(z) &=&\exp [s\sum_{m=1}^{\infty }(-1)^{m}\frac{d_{m}}{m}]\
L_{n\beta }(z)  \nonumber
\end{eqnarray}

Inverting \ref{eq:13"} by countour integration

\[
\frac{M_{n\beta }^{s}(z)}{n!}=\frac{1}{2\pi i}\oint \frac{e^{\frac{-zt}{1-t}}%
}{(1-t)^{\beta -s+1}t^{n+1}}\left( 1-2t\right) ^{-s}dz 
\]

we get an expression of \ $M$ polynomials with a useful rearrangement of
terms.In fact

expanding the power series

\[
(1-2t)^{-s}=1+2st+\frac{s(s+1)}{2!}\left( 2t\right) ^{2}+\cdots \frac{%
s(s+1)\cdots (s+k-1)}{k!}(2t)^{k}+\cdots 
\]

we get

\begin{eqnarray*}
M_{n\beta }^{s}(z) &=&L_{n(\beta -s)} \\
&&+L_{(n-1)(\beta -s)}\ 2s\left( 
\begin{array}{c}
n \\ 
1
\end{array}
\right) \cdots \\
&&+L_{(n-k)(\beta -s)}\ 2^{k}s(s+1)\cdots (s+k-1)\left( 
\begin{array}{c}
n \\ 
k
\end{array}
\right) \cdots \\
&&2^{n}s(s+1)\cdots (s+n-1)
\end{eqnarray*}

\subsection{Properties}

In the following ,unless necessary, we will use the compact notation $\exp
(s\sum )$ to mean $\exp [s\sum_{m=1}^{\infty }(-1)^{m}\frac{d_{m}}{m}]$

\begin{itemize}
\item  connection to $L$
\end{itemize}

\begin{eqnarray*}
\ \exp (-s\sum )M_{n\beta }^{s}(z) &=&L_{n\beta }(z) \\
\exp (s^{^{\prime }}\sum^{{}})M_{n\beta }^{s}(z) &=&M_{n\beta
}^{s+s^{^{\prime }}}(z)
\end{eqnarray*}

\begin{itemize}
\item  Various combinations

We may combine the $\alpha =\pm 1$ deformations and get
\end{itemize}

\begin{eqnarray*}
\exp [s\sum_{m=1}^{\infty }\left( (-1)^{m}+1\right) \frac{d_{m}}{m}]\
L_{n\beta }(z) &=&M_{n(\beta -s)}^{s}(z) \\
\exp [s\sum_{m=1}^{\infty }\left( (-1)^{m}-1\right) \frac{d_{m}}{m}]\
L_{n\beta }(z) &=&M_{n(\beta +s)}^{s}(z)
\end{eqnarray*}

\begin{itemize}
\item  Recursion formulas
\end{itemize}

\begin{eqnarray}
\frac{d}{dz}M_{n\beta }^{s}(z) &=&-M_{(n-1)(\beta +1)}^{s}(z)
\label{eq:16.5} \\
M_{n\beta }^{s^{^{\prime }}}(z) &=&M_{n(\beta -s^{^{\prime
}}+s)}^{s}+M_{(n-1)(\beta -s^{^{\prime }}+s)}^{s}\ 2(s^{^{\prime }}-s)\left( 
\begin{array}{c}
n \\ 
1
\end{array}
\right) \cdots  \nonumber \\
&&+M_{(n-k)(\beta -s^{^{\prime }}+s)}^{s}\ 2^{k}(s^{^{\prime
}}-s)(s^{^{\prime }}-s+1)\cdots (s^{^{\prime }}-s+k-1)\left( 
\begin{array}{c}
n \\ 
k
\end{array}
\right) \cdots  \nonumber \\
&&2^{n}(s^{^{\prime }}-s)(s^{^{\prime }}-s+1)\cdots (s^{^{\prime }}-s+n-1) 
\nonumber
\end{eqnarray}

The first relation is trivial .The second relation is obtained as follows

\begin{eqnarray}
M_{n\beta }^{s^{^{\prime }}}(z) &=&\exp (s\sum )\exp [(s^{^{\prime }}-s)\sum
]L_{n\beta }(z)  \nonumber \\
&=&\exp (s\sum )M_{n\beta }^{s^{^{\prime }}-s}(z)  \nonumber \\
&=&\exp (s\sum )[L_{n(\beta -s^{^{\prime }}+s)}+L_{(n-1)(\beta -s^{^{\prime
}}+s)}\ 2(s^{^{\prime }}-s)\left( 
\begin{array}{c}
n \\ 
1
\end{array}
\right) \cdots  \nonumber \\
&&+L_{(n-k)(\beta -s^{^{\prime }}+s)}\ 2^{k}(s^{^{\prime }}-s)(s^{^{\prime
}}-s+1)\cdots (s^{^{\prime }}-s+k-1)\left( 
\begin{array}{c}
n \\ 
k
\end{array}
\right) \cdots  \nonumber \\
&&2^{n}(s^{^{\prime }}-s)(s^{^{\prime }}-s+1)\cdots (s^{^{\prime }}-s+n-1) 
\nonumber
\end{eqnarray}

And applying the exponential operator on each Laguerre polynomial above we
obtain the desired result

\subsection{Differential equation}

The differential equation obeyed by Laguerre polynomials is of the form

\[
(z\frac{d^{2}}{dz^{2}}+(\beta -z+1)\frac{d}{dz}+n)L_{n}(z)=0 
\]

To find the differential equation obeyed by $M_{n\beta \alpha }^{s}(z)$ we
apply the deformation operator on both sides of the above equation and get

\begin{eqnarray*}
&&(z\frac{d^{2}}{dz^{2}}+(\beta -z+1)\frac{d}{dz}+n)M_{n\beta \alpha }^{s}(z)
\\
&=&-\left[ \exp (s\sum ),z\right] (\frac{d^{2}}{dz^{2}}-\frac{d}{dz}%
)L_{n\beta }(z) \\
&=&-s\sum_{m=1}^{n}\alpha ^{m}(\frac{d^{m+1}}{dz^{m+1}}-\frac{d^{m}}{dz^{m}}%
)M_{n\beta \alpha }^{s}(z)
\end{eqnarray*}

In the above we use the identity

\[
\lbrack \exp (s\sum_{\alpha }),z]=s\sum_{m=1}^{\infty }\alpha ^{m}\frac{%
d^{m-1}}{dz^{m-1}}\exp (s\sum_{\alpha }) 
\]

(which has a form appropriate to reproduce a differential equation ) and
limit the sum to $n$ as $M$ is identically vanishing for negative $n.$ In
the case $\alpha =1$ the right hand side reduces to $s\frac{d}{dz}M_{n\beta
(1)}^{s}(z)$ as the terms are alternating and hence compensating each other
except the first derivative ,but this is the differential equation of
Laguerre polynomials with the parameter $\beta $ shifted $\beta \rightarrow
\beta -s$ and this we know from subsection 1.3.The case of interest is $%
\alpha =-1$%
\begin{eqnarray*}
&&(z\frac{d^{2}}{dz^{2}}+(\beta +s-z+1)\frac{d}{dz}+n)M_{n\beta }^{s}(z) \\
&=&2s\sum_{m=1}^{n}(-1)^{m}\frac{d^{m}}{dz^{m}}M_{n\beta }^{s}(z) \\
&=&2s\sum_{m=1}^{n}M_{(n-m)(\beta +m)}^{s}
\end{eqnarray*}

where we use the evident recursion relation $\frac{d^{m}}{dz^{m}}M_{n\beta
}^{s}(z)=(-1)^{m}M_{(n-m)(\beta +m)}^{s}(z).\,$in \ref{eq:16.5}

This is a differential equation we can interpret equivalently as an equation
of order n with varing coefficients or a liner system of second order
differential equations or ,if we plug the explicit expression for the
polynomials $M$ into the $\sum $ , as an inhomogeneous differential equation
whose homogeneous part is the differential equation of Laguerre polynomials.

\section{ C$_{n\beta }^{s}(z)$ $,W_{n}^{s}(z)$ Polynomials}

To find the measure we assume that Polynomials $M$ are a priori partial
-orthogonal with respect to the measure $\frak{D}_{s\beta }(z)$ to be
determined, which means that we assume

\[
\int_{0}^{\infty }M_{n\beta }^{s}(z)M_{0\beta }^{s}(z)\frak{D}_{s\beta
}(z)dz\quad =\delta _{n0}\Gamma (\beta +1)\quad 
\]

\smallskip

Multiply both sides of \ref{eq:13"} by $M_{0\beta }^{s}(z)=L_{0\beta }(z)=1$
and then integrate with measure $\frak{D}_{s\beta }(z)$ we should get

\[
\int_{0}^{\infty }\frac{e^{\frac{-zt}{1-t}}}{(1-t)^{\beta -s+1}}\left(
1-2t\right) ^{-s}\frak{D}_{s\beta }(z)dz=\Gamma (\beta +1) 
\]

Let us check that the following expression of the measure is the correct one

\[
\frak{D}_{s\beta }(z)=(-1)^{s}e^{z}\frac{d^{s}}{dz^{s}}(e^{-2z}z^{\beta }) 
\]

In fact we have

\begin{eqnarray*}
&&\int_{0}^{\infty }\frac{e^{\frac{-zt}{1-t}}}{(1-t)^{\beta +1}}\left( \frac{%
1-2t}{1-t}\right) ^{-s}(-1)^{s}e^{z}\frac{d^{s}}{dz^{s}}(e^{-2z}z^{\beta })dz
\\
&=&\int_{0}^{\infty }(\frac{1-2t}{1-t})^{-s}\frac{1}{(1-t)^{\beta +1}}\frac{%
d^{s}}{dz^{s}}\left( e^{(\frac{1-2t}{1-t})z}\right) e^{-2z}z^{\beta
}dz\medskip \\
&=&\int_{0}^{\infty }\frac{1}{(1-t)^{\beta +1}}e^{\frac{-2t}{1-t}%
z}e^{-z}z^{\beta }dz\medskip \\
&=&\Gamma (\beta +1)
\end{eqnarray*}

Where the last integral is the partial -orthogonality of Laguerre
polynomials.

As for the case of $M_{n\alpha ,\text{ }H}^{s}$ polynomials( which are the
result of the action of the same deformation on Hermite polynomials ) whose
measure is $\frak{D}_{s\beta ,H}(z)=\frac{(-\alpha )^{s}}{\sqrt{\pi }}\exp
(-z^{2})H_{s}(z-\frac{\alpha }{2})$,it is possible, surprisingly , to
rewrite the measure in terms of Laguerre polynomials .After some algebra we
get

\[
\frak{D}_{s\beta }(z)=(-1)^{s}z^{\beta -s}e^{-z}L_{s(\beta -s)}(2z) 
\]

where we have used the definition of laguerre polynomials in terms of higher
derivatives $L_{n\beta }(z)=z^{-\beta }e^{z}\frac{d^{n}}{dz^{n}}(z^{\beta
+n}e^{-z})$

This is a real function of the real variable z but not positive .It is not a
measure but rather a density with no definite sign .We will however continue
to call it measure.Out of this measure ,we can form orthogonal polynomials
whose existence is related to the finitness of the integral

\[
\int_{0}^{\infty }z^{n}\frak{D}_{s\beta }(z)dz 
\]

for $n=1,2,$ $\ldots .$ . .In fact we can compute it for any value n, in
principle.To proceed we can reexpress the quantity $z^{n}\frak{D}_{s\beta
}(z)$ as a linear combination of $\frak{D}_{p\beta }(z)$ with $s-n\leq p\leq
s$

\begin{equation}
z^{n}\frak{D}_{s\beta }(z)=\sum_{p=0}^{n}\left( 
\begin{array}{c}
n \\ 
n-p
\end{array}
\right) s(s-1)\cdots (s+1-p)\frak{D}_{(s-p)(\beta +n-p)}(z)  \label{eq:48"}
\end{equation}

To see how the above expression is evaluated we worked out various
components and guessed the general pattern .Here we just show the first
component calculation

\begin{eqnarray*}
z\frak{D}_{s\beta }(z) &=&(-1)^{s}z^{\beta +1-s}e^{-z}L_{s(\beta -s)} \\
&=&(-1)^{s}z^{\beta +1-s}e^{-z}(L_{s(\beta +1-s)}-sL_{(s-1)(\beta +1-s)}) \\
&=&(-1)^{s}z^{\beta +1-s}e^{-z}L_{s(\beta +1-s)}+s(-1)^{s-1}z^{\beta
-(s-1)}e^{-z}L_{(s-1)(\beta -(s-1))} \\
&=&s\frak{D}_{(s-1)\beta }(z)+\frak{D}_{s(\beta +1)}(z)
\end{eqnarray*}

In the second line we used the known relation $L_{s\beta }=L_{s(\beta
+1)}-sL_{(s-1)(\beta +1)}.$

With the integral $\int_{0}^{\infty }\frak{D}_{s\beta }(z)dz=\Gamma (\beta
+1)$ and it is property that it does not depend on the index $s$ and formula 
\ref{eq:48"} the above integral can be computed trivially

\begin{eqnarray*}
\int_{0}^{\infty }z^{n}\frak{D}_{s\beta }(z)dz &=&\int_{0}^{\infty
}\sum_{p=0}^{n}\left( 
\begin{array}{c}
n \\ 
n-p
\end{array}
\right) s(s-1)\cdots (s+1-p)\frak{D}_{(s-p)(\beta +n-p)}(z)dz \\
&=&\sum_{p=0}^{n}\left( 
\begin{array}{c}
n \\ 
n-p
\end{array}
\right) s(s-1)\cdots (s+1-p)\Gamma (\beta +n-p+1)
\end{eqnarray*}

The most general form of $C_{n\beta }^{s}(z)$ polynomials ensuring
partial-orthogonality is a linear combination of $M_{n\beta }^{s}(z)$
polynomials i.e.

\begin{eqnarray}
C_{0\beta }^{s}(z) &=&M_{0}^{s}(z)=1  \label{eq:51} \\
C_{n\beta }^{s}(z) &=&\sum_{i=0}^{n-1}w_{i}^{n}M_{(n-i)\beta }^{s}(z) 
\nonumber
\end{eqnarray}

where the coefficients $w_{i}^{n}$ $\equiv $ $w_{p\beta }^{sn}$ are to be
determined using the orthogonality relation .We normalize polynomials $C$
above such that $w_{0}^{n}=1.$Coefficients $w_{i}^{n}$ are obtained as
follows.Define the determinant $\Delta _{n}$ and $\Delta _{n}^{^{\prime }i}$
where we use the simpler notation $M_{n}M_{m}\equiv \int_{0}^{\infty
}M_{n}(z)M_{m}(z)\frak{D}_{s\alpha }(z)dz$ and where $i$ means insertion at
the $i^{th}$ column $1\leq i\leq n$

\begin{eqnarray*}
\Delta _{n} &=&\left| 
\begin{array}{ccccc}
M_{1}M_{n} & \cdots & \cdots & M_{1}M_{2} & \text{ }M_{1}M_{1} \\ 
M_{2}M_{n} & \cdots & \cdots & M_{2}M_{2} & M_{2}M_{1} \\ 
\vdots & \cdots & \cdots & \vdots & \vdots \\ 
\vdots & \cdots & \cdots & \vdots & \vdots \\ 
M_{n}M_{n} & \cdots & \cdots & M_{n}M_{2} & M_{n}M_{1}
\end{array}
\right| \\
&&\bigskip \\
\Delta _{n}^{^{\prime }i} &=&-\left| 
\begin{array}{cccccc}
M_{1}M_{n} & \cdots & M_{n+1}M_{1} & \cdots & M_{1}M_{2} & \text{ }M_{1}M_{1}
\\ 
M_{2}M_{n} & \cdots & M_{n+1}M_{2} & \cdots & M_{2}M_{2} & M_{2}M_{1} \\ 
\vdots & \cdots &  & \cdots & \vdots & \vdots \\ 
\vdots & \cdots &  & \cdots & \vdots & \vdots \\ 
M_{n}M_{n} & \cdots & M_{n+1}M_{n} & \cdots & M_{n}M_{2} & M_{n}M_{1}
\end{array}
\right|
\end{eqnarray*}

the coefficients are then given by the formula

\[
w_{i}^{n}=\frac{\Delta _{n-1}^{^{\prime }i}}{\Delta _{n-1}} 
\]

Polynomials $W_{n\beta }^{s}(z)$ are defined as

\[
C_{n\beta }^{s}(z)=\exp (s\sum )W_{n\beta }^{s}(z) 
\]
They are the polynomials whose deformation leads to polynomials $C_{n\beta
}^{s}(z)$ .Due to formula \ref{eq:51}.They are related to Laguerre
polynomials

\[
W_{n\beta }^{s}(z)=\sum_{i=0}^{n-1}w_{i}^{n}L_{(n-i)\beta }(z) 
\]

\end{document}